# Review of canonical scenarios of gamma-ray jet emission from recent HE-VHE observations of 3C279 with MAGIC

K. Berger, J. Becerra González
*Instituto de Astrofísica de Canarias and Universidad de La Laguna, E-38206 La Laguna, Spain*

G. Bonnoli, L. Maraschi, F. Tavecchio
*INAF National Institute for Astrophysics, I-00136 Rome, Italy*

A. Domínguez
*Inst. de Astrofísica de Andalucía (CSIC), E-18080 Granada, Spain; Now at: Department of Physics and Astronomy, University of California, Riverside, CA 92521, USA*

E. Lindfors
*Tuorla Observatory, University of Turku, FI-21500 Piikkiö, Finland*

A. Stamerra
*Università di Siena, and INFN Pisa, I-53100 Siena, Italy*

On behalf of the MAGIC Collaboration

In 2006 the stand-alone MAGIC-I telescope discovered very high energy gamma-ray emission from 3C279. Additional observations were triggered when the source entered an exceptionally bright optical state in January 2007 and the Fermi space telescope measured a bright GeV gamma-ray flare in December 2008 until April 2009. While the complete January 2007 dataset does not show a significant signal, a short flare (of one day duration) has been detected on January 16th with a significance of 5.4 σ (trial-corrected). The flux corresponds to F(> 150 GeV) = (3.8 ± 0.8)·$10^{-11}$ ph cm$^{-2}$ s$^{-1}$. The December 2008 - April 2009 observations did not detect the source. We collected quasi-simultaneous data at optical and X-ray frequencies and for 2009 also γ-ray data from Fermi, which we use to determine the spectral energy distributions and the light curves. The hard γ-ray spectrum is a challenge for standard one-zone models, which are based on relativistic electrons in a jet scattering broad line region photons. We study additionally a two zone model and a lepto-hadronic model, which fit the observed spectral energy distribution more satisfactorily.

## 1. INTRODUCTION

Since autumn 2009 MAGIC is a system of two 17 m diameter imaging air Cherenkov telescopes, situated on the canary island La Palma. It is sensitive to very high energy (VHE) γ-rays between 50 GeV and several tens of TeV.

In 2006 (before the operation of the second telescope began) the single MAGIC-I telescope discovered VHE γ-ray emission from the flat spectrum radio quasar 3C 279 [1], which was the first quasar discovered to emit VHE γ-rays and is the most distant VHE γ-ray emitter known until today. It's brightness and distance across the electromagnetic spectrum make it an ideal candidate to study jet emission models as well as models of the extragalactic background light (EBL, see e.g. [2]).

## 2. DATA SAMPLE

In January 2007, 3C 279 was exceptionally bright in the optical R-band[1]. This historical high state triggered observations with the MAGIC-I telescope from January 15[th] onwards. 3C 279 was observed during nine nights in January 2007 for a total of 23.6 h, 18.6 h (seven nights) of which passed the quality selection. A VHE γ-ray flare (of one day duration) was detected on January 16th 2007 at a significance of 5.4 σ post-trial.

---

[1] http://users.utu.fi/kani/1m/3C_279_jy.html

Quasi-simultaneous data was collected at optical, IR and X-ray frequencies to determine spectral energy distributions (SED) and light curves.

During December 2008-April 2009 the source entered a bright GeV γ-ray state, as observed by the Fermi Large Area Telescopes (Fermi-LAT). Observations with MAGIC-I were triggered and (after quality selection) 11.9 h of data were collected. The main part of this dataset is from January 2009, but it also includes one night (29) in December 2008 and one (16) in April 2009. No detection in VHE γ-rays was achieved during the 2009 observation campaign. It should however be noted that the main part of the MAGIC data was taken in between the two γ-ray flares seen by Fermi.

## 3. MODELLING THE SED

In this section we discuss the SEDs of 3C 279 at three epochs: February 23, 2006, January 16, 2007 and January 21 - February 1, 2009. The VHE γ-ray data was corrected for EBL absorption using [2].

## 3.1. February 2006

We try to reproduce the 2006 data sample with a simple one-zone leptonic emission model for FSRQs (details in [3]) considering the synchrotron and inverse Compton (IC) emission from a population of relativistic electrons in a spherical emission region with radius $R$ in motion with bulk Lorentz factor Γ at an angle θ with the line of sight. The electron energy distribution is described by a smoothed broken-power law with normalization $K$ (measuring the number density of electrons at $\gamma = 1$) extending from $\gamma_1$ to $\gamma_2$, with indices $n_1$ and $n_2$ below and above the break at a $\gamma_b$, respectively. The magnetic field with intensity $B$ is supposed to be homogeneous and tangled. The seed photons for the IC emission are both the synchrotron photons produced within the jet (Synchrotron Self Compton mechanism, SSC) and those outside the jet. We consider two cases, in which the IC process occurs within or outside the Broad Line Region (BLR). In the first case the high energy emission is dominated by comptonization of the UV photons of the BLR (External Compton, EC/BLR). In the latter case the external radiation field is dominated by the IR thermal emission from the dusty torus (EC/IR, see [4] and [5]). In the case of the BLR emission we assume that the clouds are located at a distance $R_{BLR}$ from the central black hole. In the case of the torus emission, located at a distance $R_{IR}$ we assume that a fraction $\tau_{IR} = 0.5$ of the disk luminosity is intercepted and re-radiated from dust as IR emission. The result of the model fit is shown in Figure 2 and the model parameters are listed in Table 1.

The modeling of the VHE $\gamma$-ray emission from 3C 279 is particularly challenging ([1], [6]). Indeed, due to the reduced IC scattering efficiency at high energies (e.g. [7]) one zone emission models predict a soft VHE $\gamma$-ray spectrum. Both the EC/BLR and the EC/IR models can reproduce the data but require a large flux in the LAT band. Future detections of 3C 279 simultaneously in the LAT and VHE band will be crucial to confirm or rule out this possibility.

### 3.2. January 2007

At the time of the second MAGIC detection, in January 2007, 3C 279 was in a brighter optical state but in a fainter X-ray state compared to the discovery.

It is noteworthy that the VHE $\gamma$-ray flare occurred when the activity decreased in other wavebands (see Figure 1, [8]). A possible explanation could be the separation between two independent emission regions.

Motivated by the difficulties of the one-zone model discussed above (for 2007 the required GeV flux would be even higher than for 2006), we consider a "two-zone" and a lepto-hadronic model for this dataset.

In the two zone model, the emission from the optical up to the X-ray and $\gamma$-ray bands derives from a different emission region than the VHE $\gamma$-ray photons. As the VHE $\gamma$-ray flare follows the optical one, we assume that the optical up to the X-ray and $\gamma$-ray emission zone is closer to the central engine than the VHE $\gamma$-ray emission. We therefore model the spectral energy distribution assuming that the optical up to the X-ray and $\gamma$-ray bands are emitted within the BLR while the VHE $\gamma$-ray emission origins in a region outside the BLR. Assuming two regions we are doubling the number of free parameters, therefore this scenario is less constrained than one-zone models. The parameters for the two regions are reported in Table 1 and the fit to the 2007 data in Figure 3. The two-zone model can better reproduce the MAGIC data, due to the possibility to shift the peak of the EC bump to higher energies.

The lepto-hadronic model will be described in full detail in a forthcoming paper [9]. It comprises a non-thermal proton and electron distribution injected into the radiation volume as a power law with lower and upper bound ($\gamma_{min}$ and $\gamma_{max}$ respectively). The electrons may radiate through the above described mechanisms of synchrotron emission and inverse Compton scattering off synchrotron photons. The protons will also emit synchrotron radiation. Additionally they take part in photo-hadronic processes, where $\pi^0$ and $\pi^\pm$ are produced. The $\pi^0$ decays into $\gamma$-rays may eventually start a pair cascade if $\gamma$-rays are emitted in the optically thick regime. The $\pi^\pm$ decays into electrons and positrons which again emit synchrotron radiation.

In this model the number of free parameters is comparable to one-zone SSC models. We introduce four extra parameters to describe the proton population. Since we assume an identical minimum Lorentz factor and spectral index for electrons and protons, we end up with a total of nine free parameters. The fit results are reported in [10] and the model fit is shown in Figure 4.

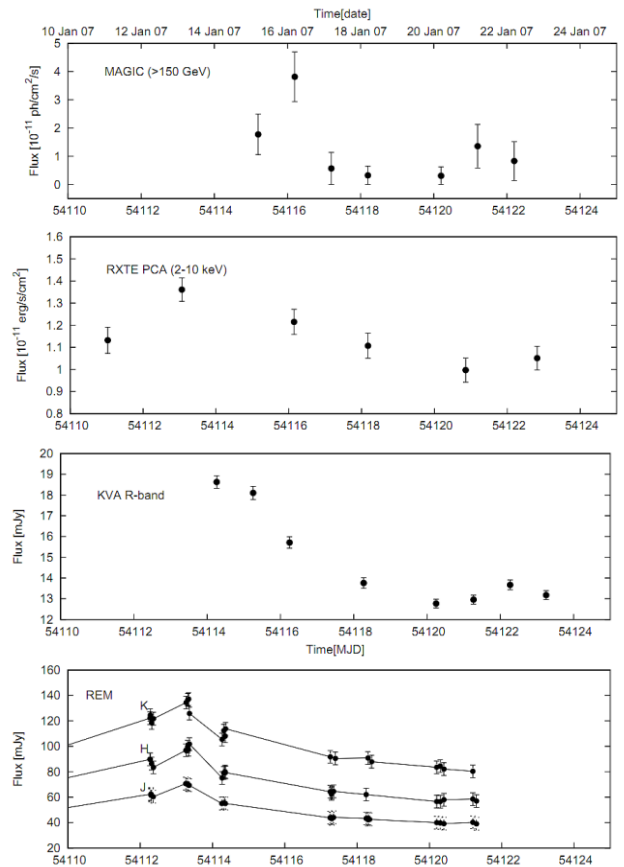

Figure 1: Light curves during the 2007 observation campaign (from top to bottom): VHE $\gamma$-ray flux >150 GeV as measured by MAGIC, RXTE PCA flux in the 2-10 keV range [8], KVA R-band observations and REM infrared observations. For the MAGIC light curve the night by night flux is calculated assuming that 3C 279 always emits $\gamma$-rays above 150 GeV.

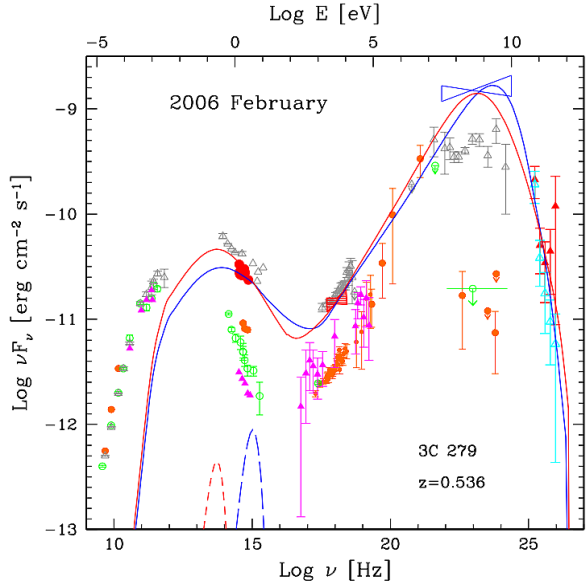

Figure 2: February 23$^{rd}$, 2006 SED of 3C 279 (red symbols: MAGIC deabsorped (triangles), RXTE (bowtie from Böttcher et al. 2009), KVA (filled circle) and optical data from literature (filled circles, Böttcher et al. 2009).. The lines assume EC emission inside (blue) and outside (red) the BLR (from a single region). Dashed lines correspond to blackbody radiation from the IR torus (red) and BLR (blue).

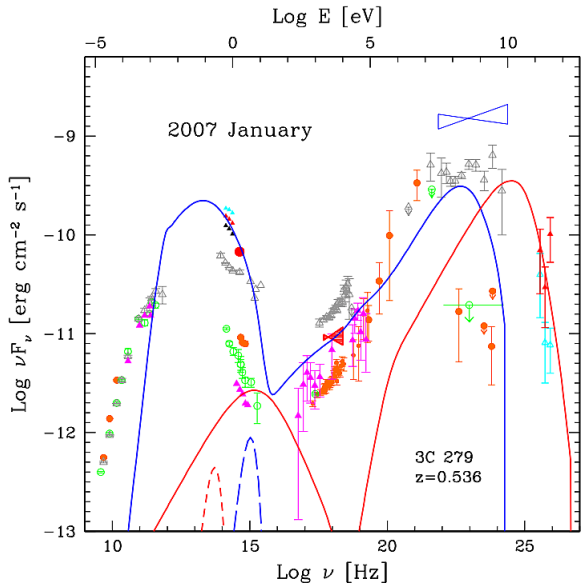

Figure 3: January 16$^{th}$, 2007 SED of 3C 279 (red symbols, like in Figure 2, but red triangles at infrared are from REM). The two lines show the emission from inside (blue) and outside (red) the broad line region. Dashed lines as in Figure 2.

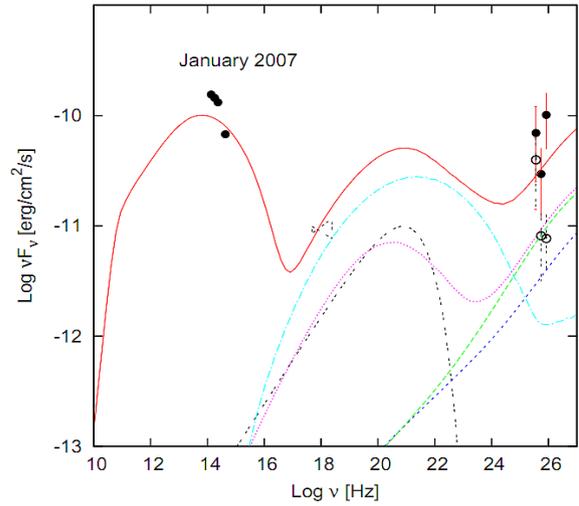

Figure 4: Lepto-hadronic model of the January 2007 SED. For a detailed list of all the components see [10]. The total fit is shown as red line.

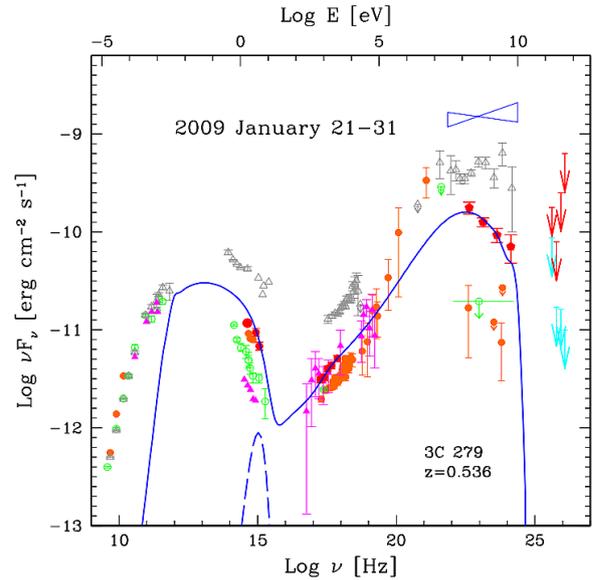

Figure 5: The spectral energy distribution of 3C 279 in January 2009 (red symbols): MAGIC (upper limits, arrows), LAT (pentagons), XRT and UVOT (red squares) and KVA (filled circle). The emission region is assumed to be inside the BLR. Blue dashed line: blackbody radiation from the IR torus.

| Model | $\gamma_{min}$ | $\gamma_b$ [$10^3$] | $\gamma_{max}$ [$10^5$] | $n_1$ | $n_2$ | $B$ [G] | $K$ [$10^4$ cm$^{-3}$] | $R$ [$10^{16}$cm] | $\delta$ | $\theta$ [deg] | $\tau_{BLR}$ | $R_{BLR}$ [$10^{17}$cm] | $R_{IR}$ [$10^{18}$cm] |
|---|---|---|---|---|---|---|---|---|---|---|---|---|---|
| 2006 One-zone BLR | 1 | 2.5 | 3.5 | 2 | 3.7 | 0.15 | 2 | 5 | 20 | 2.9 | 0.015 | 4 | – |
| 2006 One-zone IR | 1 | 2 | 2 | 2 | 4 | 0.19 | 1 | 4.5 | 27 | 2 | – | – | 4 |
| 2007 Two-zone: Opt-X-ray zone | 1 | 0.5 | 0.03 | 2 | 4.3 | 2.2 | 5.5 | 5 | 18 | 3.1 | 0.05 | 6 | – |
| 2007 Two-zone: VHE zone | 45 | 20 | 5 | 2 | 4.3 | 0.1 | 0.01 | 10 | 18 | 3.1 | – | – | 2.5 |
| 2009 One-zone BLR | 1 | 0.33 | 0.2 | 2 | 3.5 | 0.8 | 23 | 3 | 20 | 2.1 | 0.1 | 6 | – |

Table 1: Input parameters for the leptonic emission models. See text for definitions.

### 3.3. January 2009

For the 2009 observations we present a SED built using multiwavelength data nearly simultaneous to the MAGIC data taken in the period 21-31 January, which yielded in upper limits in the VHE $\gamma$-ray band. $\gamma$-ray data were derived from Fermi LAT and averaged over the same period and XRT and UVOT data of February 1, 2009 were used. The source was in a rather low state in all available bands. The SED can be modeled quite satisfactorily with a standard one zone model assuming the emission region inside the BLR (Figure 5) with typical parameters (reported in Table 1).

## 4. CONCLUSIONS

An extensive study of the multi-frequency SED and light curves of 3C 279 from the 2006, 2007 and 2009 observing seasons has been performed.

The SED of 2007 January disfavours standard one-zone SSC+EC models with an emission region inside as well as outside the BLR. A Two-zone model, wherein the VHE $\gamma$-rays are produced outside the BLR and a lepto-hadronic model can satisfactorily describe our data.

Only two more flat spectrum radio quasars have been detected in VHE $\gamma$-rays: PKS 1510-089 [11] and PKS 1222+21 [12]. In both cases the VHE $\gamma$-ray emission cannot be explained by the canonical emission scenario, as discussed in [10] and [12]. One of the possible solutions is the assumption that the VHE $\gamma$-ray emission comes from a small blob inside the jet [13].

The non-detection at VHE $\gamma$-rays during the 2009 observation period is consistent with a low optical to X-ray state and the HE $\gamma$-ray emission detected by the Fermi-LAT.

The MAGIC detection in January 2007 took place at the beginning of the rotation of the optical polarization angle, thus confirming that such events are recurrently accompanied with $\gamma$-ray flares in 3C 279 (see [10] for details).

A simultaneous detection of a $\gamma$-ray flare by MAGIC and the Fermi-LAT is necessary to test (and distinguish between) the two zone and lepto-hadronic models with better accuracy.

### Acknowledgments

We would like to thank the Instituto de Astrofísica de Canarias for the excellent working conditions at the Observatorio del Roque de los Muchachos in La Palma. The support of the German BMBF and MPG, the Italian INFN, the Swiss National Fund SNF, and the Spanish MICINN is gratefully acknowledged. This work was also supported by the Marie Curie program, by the CPAN CSD2007-00042 and MultiDark CSD2009-00064 projects of the Spanish Consolider-Ingenio 2010 programme, by grant DO02-353 of the Bulgarian NSF, by grant 127740 of the Academy of Finland, by the YIP of the Helmholtz Gemeinschaft, by the DFG Cluster of Excellence "Origin and Structure of the Universe", by the DFG Collaborative Research Centers SFB823/C4 and SFB876/C3, and by the Polish MNiSzW grant 745/N-HESSMAGIC/2010/0.